\documentclass[pra,aps,showpacs,twocolumn]{revtex4}
\usepackage{amsmath, amsthm,times}
\usepackage{amssymb,bbm}

\DeclareMathSymbol{\C}{\mathbin}{AMSb}{"43}
\DeclareSymbolFont{AMSb}{U}{msb}{m}{n}

\usepackage{graphicx}

\newcommand{\ket}[1]{|#1\rangle}
\newcommand{\id}{\mathbbm{1}}

\begin{document}

\sloppy

\title{General linear-optical quantum state generation scheme: 
\\
Applications to maximally path-entangled states}

\author{N. M.\ VanMeter$^{1}$, P.\ Lougovski$^{1}$, 
D.\ B.\ Uskov$^{1,2}$,
K.\ Kieling$^{3,4}$, J.\ Eisert$^{3,4}$, and Jonathan P.\ Dowling$^{1}$}

\address{
$^{1}$Hearne Institute for Theoretical Physics, 
Louisiana State University, Baton Rouge, LA 70803, USA \\ 
$^{2}$Department of Physics, Tulane University, 2001 Percival Stern Hall, New Orleans, LA 70118, USA \\
$^{3}$Institute for Mathematical Sciences,
Imperial College London, London SW7 2PE, UK\\
$^{4}$QOLS, Blackett Laboratory, Imperial College London, Prince Consort Road, London SW7 2BW, UK
}

\begin{abstract}
We introduce schemes for linear-optical quantum state generation. 
A quantum state generator is a device 
that prepares a desired quantum 
state using product inputs from photon sources, linear-optical networks, and post-selection using photon counters. 
We show that this device can be concisely described in terms of 
polynomial equations and unitary constraints. 
We illustrate the power of this language by applying the Gr\"{o}bner-basis technique along with the notion of vacuum 
extensions to solve the problem of how to construct a 
quantum state generator analytically for any desired state, and use methods of convex optimization to identify bounds to
success probabilities. In particular, we disprove a conjecture concerning the preparation of the maximally path-entangled 
NOON-state by providing a counterexample using these 
methods, and we derive a new upper bound on the resources 
required for NOON-state generation. 
\end{abstract}

%42.50.Lc, 03.67.-a}

\maketitle

\section{Introduction}

There are many quantum states of light, which are in great demand in quantum technology ~\cite{DowMilburn}. Due to their high robustness to decoherence, and relatively simple manipulation techniques, photons are often exploited as primary carriers of quantum information. Recently, a number of schemes have been suggested enabling quantum information processing with photons using only beam splitters, phase shifters, and photodetectors~\cite{KokReview,HwangPavelJon}.  In a different approach to quantum computing, cluster states of photons can be used to perform computation ~\cite{Raussendorf}. Moreover, exotic states of many photons, such as maximally path-entangled NOON states, have found their place in quantum metrology~\cite{Kok04}, lithography~\cite{Qlithography}, and sensing ~\cite{Kapale}. 

A natural question arises: how can these complicated states of light be prepared? One approach to the problem is to make use of an optical nonlinearity. However, due to the relatively small average number of photons involved, the overall nonlinear effect is extremely weak and typically of little practical use~\cite{GerCamp,Deb,Kow}. An alternative way to enable an effective photon-photon interaction is to use ancilla modes and projective measurements ~\cite{KokReview,HwangPavelJon}. In this way a quantum state 
generator can be realized utilizing only linear optical elements (beam splitters and phase shifters) and photon counters, at the expense of the process becoming probabilistic~\cite{KLM}. Hybrid schemes, combining weak nonlinearities and measurements, have been also proposed recently~\cite{Munro}.

Despite many theoretical efforts, the problem of quantum state preparation of light with the help of projective measurements has not been formalized and solved in a general setting. Progress with respect to the problem of constructing an optimal all-optical two-qubit gate has been achieved~\cite{Knill}. Moreover, a slightly different problem of attributing an effective physical nonlinearity to a given combination of linear-optical transformations and projective measurements has been demonstrated to have a definite answer~\cite{LapairKokSipeD}.  

In this paper we concentrate on formalizing and solving the problem of quantum state generation with linear optics and projective measurements. We first formuIate the physical problem of state generation in terms of polynomial equations in Section II. We then argue that the equations obtained can be solved analytically, by applying standard tools of algebraic geometry, when a unitarity constraint is relaxed. Moreover, we show how unitarity can be reintroduced into the solution by addition of auxiliary vacuum modes. In Section III we illustrate the formalism by considering an example of a ``NOON" state generator. We demonstrate how an optimal ``NOON" state generator can be constructed using convex optimization tools. A simple example for 5-photon ``NOON" state generation disproves the ``No-Go" conjecture~\cite{HwangPavelJon}. We draw conclusions and summarize in Section IV. 

\section{Linear-optical quantum state generator}
Any linear-optical quantum state generator (LOQSG)
can be thought of as consisting of two main blocks (see Fig. 1). 
The first block is a $N$-port linear-optical device, described by a unitary matrix $U\in U(N)$
combining $N$ input into $N$ output modes. 
The second block represents a projective measurement 
of some of the modes of the first block, in which a certain 
pattern of photons measured in some $M<N$ of the modes is considered a ``successful measurement'', leading to a preparation of the
desired state in the remaining modes (compare also 
Refs.~\cite{KLM,Knill,Prep1,Prep2}). This measurement is
probabilistic --- but heralded --- or ``event-ready''.
Clearly, the output is determined by an interplay between the numbers of input photons, the entries of the matrix $U$, and the numbers of photons detected.

There are two types of problems that can be formulated around the concept of LOQSG. The first problem is the following: 
Given the matrix $U$  and a known input state, which output states can be generated for different projective
measurements? This is what could be referred to as the 
``forward problem''. 
This question is equivalent to the problem of finding the effective nonlinearity generated by a given projective measurement 
and was addressed in Refs.~\cite{LapairKokSipeD,Scheel}. 
The second problem is that of state preparation:
Given an input state, a projective
measurement, and a target state, is it possible to 
determine a unitary matrix $U$, of appropriate dimension, 
involving potentially further auxiliary modes, 
describing the unitary block of the LOQSG? 
Given a certain input state, this important problem asks whether a device for the 
preparation of a certain quantum state
can be identified, and if so, what elements it contains.  Once the unitary is found, there is a simple, well-known prescription for converting it to an optical implementation with beam-splitters and phase-shifters, etc. ~\cite{ReckZeilinger}.
Finding such a unitary we call the ``inverse problem''.
In this section, we provide a mathematical description of 
LOQSG and illustrate how methods of algebraic geometry can be used 
to solve this inverse problem.

\begin{figure}
  \includegraphics[width=6cm]{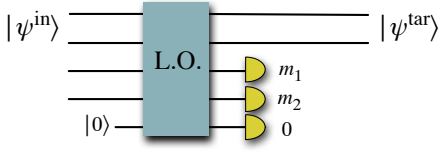}
  \caption{The linear optical quantum state generation scheme.}
\end{figure}

To be more specific, we start from a given input
$|\psi^{\text{in}}\rangle$ 
in $N$ optical modes obtained from photon sources. 
Then, we investigate all state vectors that can be reached
from this using arbitrary networks of linear 
optical elements~\cite{ReckZeilinger}. We hence investigate the orbit 
$\Omega=\{|\Psi\rangle :  |\Psi\rangle = \mathcal{U}(U)
|\psi^{\text{in}}\rangle\,\text{ for some } 
U \in U(N)\}$ of an input state
vector $|\psi^{\text{in}}\rangle$. The unitary 
$\mathcal{U}(U)$ acting in the Hilbert space of quantum states 
is the standard Fock-Bargmann representation
of the unitary transformation $U$ acting on modes.
If we now denote with $|\psi^{\text{tar}}\rangle$ 
the desired output in $N$-$M$ of the modes upon a particular projective measurement 
 $P=|\psi^{\text D}\rangle\langle \psi^{\text D}| $ on $M$ of the modes, 
then 
\begin{eqnarray*}
	\Theta=\{|\Psi\rangle : |\Psi\rangle = 
	\alpha |\psi^{\text{tar}}\rangle|\psi^{D}\rangle + \sum_{i}\beta_{i} |	\psi^{\text{any}}_{i} \rangle|\psi^{D\bot}_{i}\rangle,\alpha > 0\} 
\end{eqnarray*}
is the set of all state vectors which can be converted 
into $|\psi^{\text{tar}}\rangle$ by the measurement of 
$|\psi^{\text D}\rangle$ in auxiliary modes. The $|\alpha|^2$ 
represents a probability of success,
and each $|\psi^{\text D\bot}_i\rangle$ denotes a state
vector orthogonal to $|\psi^{\text D}\rangle$.  The solution to the state preparation problem is then the intersection of $\Omega$ and $\Theta$.

Since there exists a one-to-one correspondence between photonic states and (possibly infinite) polynomials in the creation operators of 
the modes~\cite{Perelomov}, the problem of 
finding the intersection of $\Omega$ and $\Theta$ can be recast in terms of polynomial equalities.  We will assume throughout the paper that the input has been prepared using individual sources, so the input states are product states of Fock states with respect to $N$ modes: $|\psi^{\text{in}}\rangle = |n_{1},\dots , n_{N}\rangle$. These input modes are associated with
$N$ creation operators $a^\dagger = (a_1^\dagger,\dots,a_N^\dagger)$.
The mode transformation of the LOQSG
is given by a $N\times N$ unitary matrix $U$. 
In other words, the creation operators transform as
$(a^{\dagger})^T = U(a^{\dagger}_{\text{out}})^T$, 
where $a^{\dagger}_{\text{out}}= (a^{\dagger}_{1,\text{out}},\dots,
a^{\dagger}_{N,\text{out}})$ denotes the creation operators of 
the output modes. The output state of the LOQSG before a projective 
measurement hence reads
\begin{eqnarray}
	|\psi^{\text{out}}\rangle & = &
F(a^{\dagger}_{1,\text{out}},\dots,a^{\dagger}_{N,\text{out}})|0\rangle \nonumber \\ &:=&
	\prod\limits^{N}_{i=1} \frac{1}{\sqrt{n_{i}!}}
	\biggl(\sum\limits^{N}_{j=1}U_{i,j}
	a^{\dagger}_{j, \text{out}}
	\biggr)^{n_{i}} |0\rangle . \label{outstate}
\end{eqnarray}
Here, $F(a^{\dagger}_{1,\text{out}},\dots,a^{\dagger}_{N,\text{out}})$
is a homogeneous polynomial
of degree $n=\sum^{N}_{k=1}n_{k}$ in the creation operators.
Coefficients of the monomials in $F$ 
are in turn homogeneous polynomials of degree $n$ 
belonging to the polynomial ring $\C [U_{i,j}]$. 

The output of the LOQSG after a successful projective measurement of $(m_1,\ldots,m_M)$ photons in modes $N-M+1,\dots, N$ is then given by 
\begin{equation*}
|\Phi\rangle =\langle m_1,\dots,m_M  |\psi^{\text{out}}\rangle.
\end{equation*} 
The state vector
$|\Phi\rangle$ is the result of the action of a 
polynomial $G(a^{\dagger}_{1,\text{out}},\dots,a^{\dagger}_{N-M,\text{out}})$ on the vacuum $|0\rangle$. 
The polynomial $G$ is constructed from $F$ by selecting all terms which contain $a^{\dagger}_{N-M+1,\text{out}},\dots,a^{\dagger}_{N,\text{out}}$ exactly
to the powers $m_1,\dots,m_M$, respectively, and replacing these creation operators by $1$ afterwards.
Thus, the polynomial $G$ acting on vacuum describes the final state in the first $N$-$M$ unmeasured modes, which form
the output of the device. 

For further convenience, we have summarized all symbols used in our paper relevant to our description of the LOQSG and their corresponding meaning in the table contained in Appendix A.

By construction $G$ is a homogeneous polynomial of degree $n-m$, $m=\sum_{k=1}^{M}m_{k}$, with coefficients in $\C [U_{i,j}]$.
Similarly, the desired output of the LOQSG, $|\psi^{\text{tar}}\rangle$, 
is a state vector in the modes $1,\dots, N$-$M$, and can be written as a polynomial $Q(a^{\dagger}_{1,\text{out}},\dots,a^{\dagger}_{N-M,\text{out}})$ acting on vacuum. 
The problem of finding a LOQSG is then equivalent to 
finding a unitary matrix $U$ such that
$G= \alpha Q$, for some nonzero $\alpha$.

%For further convenience we have summarized all elements of mathematical description of LOQSG in the Table I.

We notice that for the polynomial equality $G=\alpha Q$ to hold true, 
the coefficients of like monomials in $G$ and $\alpha Q$ must be equal.  
This leads to a system of polynomial equations in the 
variables $U_{i,j}$, the entries of our 
transformation matrix $U$.  This system of polynomial equations, 
along with the constraint that $U$ be unitary, is the mathematical 
formulation of the state preparation problem.  In general,
there is no efficient solution technique for solving such a system. It is because the unitarity constraint involves the operation of complex conjugation and hence cannot be written in an algebraic form, i.e., 
in the same way as the polynomial equations. So, we instead propose a method for finding the mode transformation matrix for a larger LOQSG described by input $(n_1,\ldots,n_N,0,\ldots,0)$ and measurement $(m_1,\ldots,m_M,0,\ldots,0)$.  The method involves two steps:  (1) For the original system, we relax the constraint that the matrix $U$ be unitary, and solve the polynomial system $G=\alpha Q$ for non-unitary matrix $A$ with entries $A_{i,j}$; (2) For $A$ as found in step (1), we find a larger matrix $U'$ which is unitary and contains $A$ as a submatrix; this $U'$ describes the LOQSG, and will generate the desired state $\ket{\psi^{\text{tar}}}$ in the unmeasured modes.

To realize the step (1) and to solve the polynomial system $G=\alpha Q$, we use the Gr\"obner basis technique~\cite{CLO}. 
We first find a Gr\"obner basis, $\{g_1,\dots,g_l\}\subset \C [A_{i,j}]$, for the ideal generated by the coefficients of $G-\alpha Q$.
This Gr\"obner basis is the minimal set of polynomials that do not leave a remainder when the original polynomials are divided by them. One of the essential features of this approach is contained in the elimination theorem,
which states that the Gr{\"o}bner basis consists of polynomials of only the first variable in the ordering, 
\begin{equation*}
g_1=g_1(A_{1,1})\subset\C[A_{1,1}], 
\end{equation*}
the first two variables, 
$g_2=g_2(A_{1,1},A_{1,2})\subset\C[A_{1,1},A_{1,2}]$,
and so forth. This property allows us to simply
find the solution to the whole system by solving for each
variable subsequently --- a procedure that is often regarded
as extension of a solution. That means finding the solutions $A^0_{1,1}$ of $g_1(A_{1,1})=0$,
$A^0_{1,2}$ of $g_2(A^0_{1,1},A_{1,2})=0$, and so on. Thus, the whole problem is reduced
to finding roots of monovariate polynomials.  In this way, the Gr\"obner basis technique is the generalization to polynomial systems of the technique of Gaussian elimination for linear systems, and is simple to implement in algebraic software.

For the case of product-state inputs, we can simplify the polynomial system of equations by eliminating some arbitrary factors before using the Gr\"obner basis technique.  We remark that for any matrix $A$, which is a solution to the polynomial system, a rescaling of the rows of $A$ --- corresponding to the input modes of the product state --- by arbitrary nonzero factors $x_i$ and of the columns of $A$ --- corresponding to the measured modes --- by arbitrary nonzero factors $y_j$ will generate another solution to the system.  Hence, for any solution $A$, we define the equivalence class of $A$ by 
$[A]=\{XAY : X=\text{diag}(x_1,\ldots,x_N),Y=\text{diag}(y_1,\ldots,y_N),y_1=\ldots=y_{N-M}=1\}$.  In any such equivalence class, there is exactly one member $A$ for which $A_{1,1}=\ldots = A_{N,1}=A_{1,N-M+1}=\ldots=A_{1,N}=1$.  Hence, by setting the aforementioned variables to 1 from the very beginning, one can solve a much simpler polynomial system in the remaining variables using the Gr\"obner basis technique; each solution of that system will correspond to an equivalence class of solutions for the entire system.  We call such a solution an equivalence class representative.

For a given input and projective measurement, one will find, in general, many equivalence classes of solutions.  In most cases, an equivalence class representative $A$ will not be unitary and hence will not conserve the canonical commutation relations, which implies that $A$ and hence an underlying LOQSG is not physically realizable. However, for a non-unitary $A$, one can think of a larger optical system described by a unitary matrix $U'$, and hence experimentally implementable, in $N'=N+d$ modes which in turn contains a member of $[A]$ as a submatrix.  Physically, this corresponds to an addition of auxiliary modes to the LOQSG.  Then, if we input and measure vacuum in these additional modes, the entries of $U$ corresponding to the added modes do not affect the final state of the original modes.  Given an input on $N$ modes and a desired output on $N-M$ modes, the (i) total number of modes $N'$, the (ii) unitary $U'\in U(N')$, and the (iii) measurement pattern $(m_1,\dots , m_M,0,\dots, 0)$ reflecting success are then the solution to the problem of finding a LOQSG. The question arises: for a given non-unitary matrix $A$, when is it possible to find a matrix $U'$ which is unitary and contains some member of $[A]$ as a submatrix? Moreover, how can we find the extension which optimizes the success probability of the LOQSG?

It is not difficult to see that any matrix $A$ can be extended if 
$\lambda=|||A|||\le1$, so if the largest singular value of 
$A$ is not larger than unity (compare also Ref.\ \cite{Knill}). 
Hence for any $A$, we can always extend $A/\lambda \in [A]$ to a unitary.  Let the singular value decomposition of $A$ be denoted as 
%\begin{equation}
	$A=VDW$, 
%\end{equation}
where $V,W \in U(N)$ are unitary and $D=\mbox{diag}(d_1,\ldots,d_N)$.  Then, the minimum dimension of the extended unitary $U$ containing $A/\lambda$ is $N+d$, 
where $d=\mbox{rank}(D/\lambda-\id)$. Thus, $d<N$ additional vacuum modes are sufficient to extend a member of $[A]$ to a unitary.  The minimum number of additional modes needed to extend at least one member of $[A]$ remains unsolved. 
However, we can solve the problem of optimizing the success probability of the extended LOQSG using, for example, ideas from optimization theory~\cite{Eisert}.

For any solution of the polynomial equations the success probability of correct state preparation is by definition given by $|\alpha|^2$.
The allowed rescaling of rows and columns rescale the amplitudes in the respective input and output modes such that within an equivalence class
\begin{equation}
	 p_{\mathrm{s}} =
	\left| \alpha \prod_k x_k^{n_k} \prod_l y_l^{m_l} \right|^2,
  \label{eqn:ps}
\end{equation}
where $p_{\mathrm{s}}$ is the success probability and $x_k$ and $y_l$ are the arbitrary row and column multipliers as presented in the definition of our equivalence classes.
Note, that w.l.o.g. it is sufficient to chose $x_k$ and $y_l$ real and positive, thus every quantity in (\ref{eqn:ps}) is real.
The problem is then to maximize $p_{\mathrm{s}}$ subject to the constraint 
%\begin{equation*}
	$XAY^2 A^\dagger X\leq \id\nonumber$.
%\end{equation*}	
For simplicity we will focus on the important case when $n_i=n_j$ for all $i,j=1,\dots ,N$ and $m_i=m_j$ for all $i,j=1,\ldots,M$. Then, for given $Y$, the constraint can be written as a
so-called semi-definite constraint. Semi-definite optimization problems can be efficiently solved and solvers are readily available \cite{Semi}.  As $A Y^2 A^\dagger$ is invertible, the constraint is equivalent to 
	\begin{eqnarray*}
	\left[
	\begin{array}{cc}
	\id & X\\
	X & (A Y^2 A^\dagger)^{-1}
	\end{array}		
	\right]\geq 0. 
	\end{eqnarray*}
	The objective function,
	$p_{\mathrm{s}}=c (\prod_k x_k)^{2n_1}$ for
a constant $c=\alpha^2(\prod_ly_l)^{2m_M}$, is a monomial, and hence clearly not linear. However, this monomial 
can be relaxed to again a hierarchy of semi-definite constraints, without altering the optimal objective value:
Let $s$ be the smallest integer such that $2^s \geq N$.
Let us assume that $2^s = N$; if
$N$ is smaller, we can always pad with variables that we enforce
to be unity by means of linear constraints. 
Then, let $x^{(1)}_k = x_k$ for $k=1,\dots, N$, and
$	x_{k-1}^{(j)} x_k^{(j)}= (x_{k/2}^{(j+1)} )^2$ for 
$k\in \{2,4,\dots, 2^{s+1-j} \}$ and $j=1,\dots, s$. Hence, we 
have introduced a number of new variables, according to a 
hierarchy. Each of the quadratic equality constraints 
of this form can actually be enforced,
as is easy to see, see footnote~\cite{Opt}. In fact, given $Y$,
we have written the above problem as a semi-definite
problem. In turn, relaxations give efficient upper bounds to the success probability
for the  LOQSG when simultaneously varying $X$ and $Y$.  The solution to this problem gives rise to the LOQSG 
operating with the highest probability of success within the equivalence class of $A$. Then, for a given input and measurement, the overall optimal LOQSG can be found by simply optimizing within each equivalence class by the method described above, and choosing the best equivalence class.

\section{Example: NOON-state LOQSG}

To illustrate how to solve the problem of the intersection of $\Omega$ and $\Theta$ using the language of polynomials, let us consider the state generation problem for target states of the form 
\begin{equation*}
	\ket{\psi^{\text{tar}}}=(\ket{n,0}+\ket{0,n})/\sqrt{2},
\end{equation*} 
the so-called path-entangled NOON state~\cite{LeeKokDowling}. The NOON state is important in a number of applications such as quantum metrology and lithography~\cite{Qlithography}. It cannot, however, be generated using only linear optics from product sources, 
and, therefore, the NOON-state generation problem is both important and nontrivial. 
For NOON-state generation, we require
\begin{equation}
	\label{reducible}
	\mbox{Coefficient}[F,\prod_{j=3}^N (a_{j,\text{out}}^{\dag})^{m_{j}}]=\frac{\alpha}{\sqrt{2 n!}} ((a_{1,\text{out}}^{\dag})^n+(a_{2,\text{out}}^{\dag})^n).
\end{equation}
Eq. (\ref{reducible}) cannot be satisfied for input states for which $n_i > m+1$ for any $i \in \{1,\ldots,N\}$.  To see this, note that the RHS of Eq. (\ref{reducible}) has a unique factorization into distinct linear terms.  If $n_i>m+1$ for any mode $i$ the LHS would necessarily have multiple identical linear factors of in $a_{1,\text{out}}^{\dag}$ and $a_{2,\text{out}}^{\dag}$.  Hence, for the equality to hold, we must have $n_i \leq m+1$. Thus, using polynomial properties, we have shown that the intersection of $\Omega$ and $\Theta$ is empty for any $\ket{\psi ^{\text{in}}}$ which contains more than $m+1$ photons in any of its modes.  In other words, we have demonstrated that the largest NOON state that might be generated from $N$ modes with a total of $m$ measured photons is $N(m+1)-m$.

The argument above is only useful for determining whether an intersection is empty. 
However, the ultimate goal of this section is to present a constructive method to find all points in the intersection when it is non-empty.  
To illustrate our method, we demonstrate how to build a NOON-state LOQSG.  Let us consider the following LOQSG input $\ket{\psi^{\text{in}}}= \ket{2,2,2}$.  In this case the mode transformation matrix $A$ is a $3\times 3$ matrix with $9$ complex entries $A_{i,j}$. 
We are particularly interested in a projective measurement of one photon in mode $3$, 
i.\,e., $|\psi^{\text D}\rangle=|m_1\rangle = |1\rangle$. 
The total number of photons in the remaining 
(unmeasured) modes is thus $5$ and, 
therefore, the only NOON state that can be created is 
the one corresponding to the state vector
$\ket{\psi^{\text{tar}}}= (\ket{5,0}+\ket{0,5})/\sqrt{2}$. 
Proceeding as discussed in Section II we calculate the polynomials $F$, $Q$, and $G$. They read,
\begin{eqnarray}\label{polys}
  	F & = & \frac{1}{\sqrt{8}}\prod_{i=1}^{3}
	\left(\sum_{j=1}^{3}A_{i,j}
	a^{\dagger}_{j,\text{out}}\right)^2, \nonumber\\ 
  	G & = & \mathrm{Coefficient}
	[F(a^{\dagger}_{1,\text{out}},a^{\dagger}_{2,\text{out}},a^	
	{\dagger}_{3,\text{out}}), a^{\dagger}_{3,\text{out}}], \\
  	Q & = & \frac{1}{\sqrt{2\cdot5!}}
	\left((a^{\dagger}_{1,\text{out}})^5 + 
	(a^{\dagger}_{2,\text{out}})^5\right). \nonumber
\end{eqnarray}
The equality $G=\alpha Q$ leads to a system of six polynomial equations for the coefficients of 
$(a^{\dagger}_{1,\text{out}})^k(a^{\dagger}_{2,\text{out}})^{5-k}$, 
$k=0,\dots,5$,
with respect to ten complex variables: the nine elements of $A$ and $\alpha$.

We now simplify the system to solve only for equivalence classes by setting $A_{1,1}=A_{2,1}=A_{3,1}=A_{1,3}=1$.  Once the Gr\"obner basis technique is used to solve the simplified system, one finds that there are finitely many equivalence classes.  Here, we will illustratively show the work for one of them (in fact, for an optimal one).  The following matrix is the optimal equivalence class representative:
	\begin{equation}\label{3x3matrix}
		A = \left[
        \begin{array}{ccc}
            1 & 1 & 1 \\
            1 & e^{-\frac{4\pi i}{5}} & \frac{3+\sqrt{5}}{2}e^{-\frac{2\pi i}{5}} \\
            1 & e^{\frac{4\pi i}{5}} & \frac{3+\sqrt{5}}{2}e^{\frac{2\pi i}{5}}
        \end{array}
        \right] 
        \end{equation}
Although the mode transformation in Eq.\ 
(\ref{3x3matrix}) generates the final desired NOON state 
vector up to normalization
\begin{equation*}
	\alpha (|5,0\rangle + |0,5\rangle)/\sqrt{2}
\end{equation*}	
with $\alpha = \sqrt{30} (3+\sqrt{5})$, it
is easy to see that the matrix $A$ is non-unitary and must be extended.

We now seek to optimize the success probability, $p_{\mathrm{s}} = \alpha^2 (x_1 x_2 x_3)^4 (y_3)^2$ of the extended unitary $U$ over all $x_1,x_2,x_3$, and $y_3$ such that $XAY$ is extendable.  Let us view $x_1,x_2,x_3$ in 
spherical coordinates:  
$p_{\mathrm{s}}=\alpha^2 r^{12} (\sin^2\theta \cos\theta \sin \phi \cos\phi)^4 y_3^2$.  
For a given direction $\theta,\phi$, the optimal $p_{\mathrm{s}}$ is 
reached when $r$ is chosen maximally; thus, we choose $r$ so that the largest singular value of $XAY$ is 1. For a given $y_3$, then, the optimal success probability of the extension over all $X$ can be found by varying $\theta$ and $\phi$ over an entire sphere; call this $p_{\mathrm{s,opt}}$.  Then, by evaluating $p_{\mathrm{s,opt}}$ over a reasonably large range of $y_3$, the global optimum of the extension can be found.

Using the optimization procedure described above, we find that the global maximal success probability is reached at a point $(x_1,x_2,x_3,y_3)$ for which the matrix can be extended minimally, i.e., by only 1 additional dimension.  It remains unclear whether this result will hold true in general.  One can check that the unitary matrix  
\begin{displaymath}\label{4x4unitary}
	 \left[
	\begin{array}{cccc}
	0.5722 & 0.5722 & 0.1894 & 0.5561 \\
	0.5257 & 0.5257e^{-\frac{4\pi i}{5}} & 0.4556e^{-\frac{2\pi i}{5}} & 0.4895e^{\frac{3\pi i}{5}} \\
	0.5257 & 0.5257e^{\frac{4\pi i}{5}} & 0.4556e^{\frac{2\pi i}{5}} & 0.4895e^{-\frac{3\pi i}{5}} \\
	0.3461e^{\pi i} & 0.3461e^{\pi i} & 0.7409 & 0.4599
	\end{array}
	\right]
\end{displaymath}
contains a member of $[A]$ as a submatrix and generates a unitary transformation on four modes which results in a 
five-photon NOON state starting with input $\ket{2,2,2,0}$ and projective measurement of one photon in mode $3$ and vacuum in mode $4$ with optimal success probability $\approx 0.05639$.  By using the well-known algorithm presented in Ref.~\cite{ReckZeilinger}, the above unitary matrix can easily be transformed into a linear-optical network consisting of phase-shifters and beam-splitters and, as such, can be constructed in the lab.

Remarkably, the example we have just considered also serves as a counterexample to the ``No-Go'' conjecture~\cite{HwangPavelJon}. The conjecture states that the largest NOON state that can be produced from product inputs with only $N$ modes is that of $N$ photons. However, we have seen that there is a situation when the NOON state of five photons can be generated in only four modes. 

\section{Conclusions and Outlook}

In this work, we have shown how the problem of identifying a
linear optical state preparation device 
can be formulated and solved using the language of polynomials. In this way, we do not have to include unitarity conditions as constraints on our system. Instead, we solve the polynomial equations using the methods of algebraic geometry and later restore unitarity with a sacrifice of at most doubling the number of modes.  We have 
introduced a general framework that allows for the systematic
construction of linear optical devices preparing 
entangled quantum states of light.

It is worth noting that in addition to solving the problem of state generation by linear-optical means, the technique we propose can also be used to construct optimal linear-optical quantum gates. With care taken to formulate appropriate equivalence classes, the problem of constructing a (probabilistic) linear-optical quantum gate involves the same system of equations, and hence can be solved using Gr\"obner bases and vacuum extensions. Applied to gate construction, our solution scheme can be seen as the generalization of the procedure used to construct an optimal linear-optical NS gate in Ref.\ \cite{Knill}.

\section{Appendix A}

\begin{tabular}{|l|l|}
\hline
\textbf{Symbol} & \textbf{Meaning}
\\
\hline
$N$ & total number of modes\\
$M$ & number of ancillary modes\\
$U$ & $N$--dim. unitary representing the linear optics\\
$n_i$ & number of photons input in mode $i$\\
$n$ & $\sum_{i=1}^N n_i$ -- total number of input photons\\
$m_i$ & number of photons measured in mode $N-M+i$\\
$m$ & $\sum_{i=N-M+1}^N m_i$ -- detected photons number\\
$\ket{\psi^{\text{in}}}$ & $N$--dim. input product state\\
$\ket{\psi^{\text{out}}}$ & $N$--dim. state after linear optics\\
$\ket{\psi^{\text{D}}}$ & $M$--dim. specified measurement state\\
$\ket{\Phi}$ & $N-M$--dim. state remaining after measurement\\
$\ket{\psi^{\text{tar}}}$ & $N-M$--dim. desired output state\\
$F$ & polynomial representing $\ket{\psi^{\text{out}}}$\\
$G$ & polynomial representing $\ket{\Phi}$\\
$Q$ & polynomial representing $\ket{\psi^{\text{tar}}}$\\
$\alpha$ & probability amplitude for successful measurement\\
$A$ & $N$--dim. matrix solution to $G=\alpha Q$\\
$d$ & number of vacuum modes added to system\\
$N'$ & $N+d$\\
$U'$ & $N'$--dim. unitary containing $A$ as a submatrix\\
\hline
\end{tabular}

\section{Aknowledgement}

NMV, PL, DU, and JPD acknowledge support from the 
the Army Research Office
and the Disruptive Technologies Office. 
KK and JE acknowledge support 
from the DFG (SPP 1116), the EU (QAP), 
the EPSRC, the QIP-IRC, the 
EURYI Award, and Microsoft Research through the European PhD Scholarship Programme.


\begin{thebibliography}{99}
\bibitem{Opt}
	To start with, we can maximize $t$, subject to 	
	\begin{equation*}
	\biggl[
	\begin{array}{cc}
	x_1^{(s)} & t\\
	t & x_2^{(s)}
	\end{array}
	\biggr]\geq 0.
	\end{equation*}
	 Then, each of the equality constraints
	can be enforced by the hierarchy
	\begin{equation*}
	\biggl[
	\begin{array}{cc}
	x_{k-1}^{(j)} & x_{k/2}^{(j+1)}\\
	x_{k/2}^{(j+1)} & x_k^{(j)}
	\end{array}
	\biggr]\geq 0
	\end{equation*}
	for  $k\in \{2,4,\dots, 2^{s+1-j} \} $
	and $j=1,\dots, s$.
\end{thebibliography}
\end{document}